\documentclass[letterpaper, 10 pt, conference]{ieeeconf}
\IEEEoverridecommandlockouts

\usepackage{graphicx}
\usepackage{caption}
\usepackage{algpseudocode}
\usepackage{amsmath}
\usepackage{longtable}
\usepackage{adjustbox}

\usepackage{amsmath,amssymb,amsfonts,epsf,epsfig,times}
\usepackage[all]{xy}
\usepackage{graphicx,color}
\usepackage{subfigure}
\usepackage{url}
\usepackage{cite}
\usepackage{tikz}
\usepackage{epstopdf}
\usepackage{algorithm}
\usepackage[]{amssymb}

\def\QED{\mbox{\rule[0pt]{1.5ex}{1.5ex}}}

\newcommand{\abs}[1]{\left|#1\right|}

\newcommand{\OMIT}[1]{}

\makeatletter
\def\BState{\State\hskip-\ALG@thistlm}
\makeatother

\begin{document}
\title{A Multi-Layer K-means Approach for Multi-Sensor Data Pattern Recognition in Multi-Target Localization}
\author{Samuel Silva, Rengan Suresh, Feng Tao, Johnathan Votion, Yongcan Cao
\thanks{The authors are with the Department of Electrical and Computer Engineering, The University of Texas, San Antonio, TX 78249. 
}
\thanks{Corresponding Author: Yongcan Cao (yongcan.cao@utsa.edu)}
}

\markboth{}
         {}

\maketitle

\begin{abstract}

Data-target association is an important step in multi-target localization for the intelligent operation of unmanned systems in numerous applications such as search and rescue, traffic management and surveillance. The objective of this paper is to present an innovative data association learning approach named multi-layer K-means (MLKM) based on leveraging the advantages of some existing machine learning approaches, including K-means, K-means++, and deep neural networks. To enable the accurate data association from different sensors for efficient target localization, MLKM relies on the clustering capabilities of K-means++ structured in a multi-layer framework with the error correction feature that is motivated by the backpropogation that is well-known in deep learning research. To show the effectiveness of the MLKM method, numerous simulation examples are conducted to compare its performance with K-means, K-means++, and deep neural networks. %
\end{abstract}


\IEEEpeerreviewmaketitle


\section{Introduction}

Multi-target tracking (MTT) is focused on the accurate detection and localization for multiple dynamic targets when measurements from these targets often come from numerous spatially distributed sensors. Obtaining the locations of the targets can be complex when sensors have limited sensing capabilities. Due to the potential applications of MTT, MTT can be dated back to 1960's initially related to aerospace applications  \cite{vo2015multitarget}. The theoretical advances in MTT, new sensor capabilities, and more computational power have made it possible to apply MTT in numerous applications such as surveillance \cite{benfold2011stable,haritaoglu2000w}, computer vision \cite{ullah2016hog, kamal2016distributed}, network and computer security \cite{zhang2008global} and sensor network \cite{vercauteren2005joint}.
	
In general, solving the MTT problem involves three tasks: \textit{(i) Extraction} -  extract target related information from the raw data acquired from the sensors; \textit{(ii) Data association} - identify each target's corresponding measurements; and, \textit{(iii) Estimation} - estimate the position of targets via single target tracking techniques (as shown \cite{liu1994single,de2010augmentation,goyal2015implementation}). Perhaps the most challenging task is to conduct data association because if data associated with each target is determined, it becomes much easier to conduct estimation for each individual target. In this paper, our focus is also on the data association problem.
	
The main objective of this paper is to investigate the applicability of machine learning algorithms for the data association problem and then develop a new multi-layer learning algorithm by leveraging the advantages of different machine learning algorithms. Extracting data patterns from spatially distributed sensors is crucial to obtain accurate knowledge about the underlying activities in numerous applications, such as disaster response~\cite{GeorgeEtAl10}, border patrol~\cite{OnurEDA07}, force protection~\cite{AhmedCCK15,CasbeerMeierCao13}, combat missions~\cite{BokarevaEtAl06}, and traffic management~\cite{min2011real}.
	
	
The perhaps most popular algorithms for MTT are based on probabilistic reasoning techniques, including Joint Probabilistic Data Association Filter (JPDAF) \cite{bar2011tracking, ma2006distributive,kim2016jpdas,yuhuan2014modified}, Multiple Hypothesis Tracking (MHT) \cite{Blackman04}, and Random Finite Set (RFS) \cite{mahler2014advances}. The JPDAF \cite{fortmann1983sonar} is built on the probability data association filter (PDAF), which applies Bayesian filter to the average of all measurements weighted according to their association probabilities in order to perform MTT \cite{vo2015multitarget}. One problem with the JPDAF is that the computation complexity grows exponentially as the number of targets and measurements increase. In \cite{musicki2004joint,musicki2008multi,bar1989automatic}, modifications to deal with exponential complexity have been proposed, however at the cost of accuracy. The MHT \cite{reid1979algorithm, zhao2014novel, jensfelt2001active, leibe2007coupled} is a multi-scan tracking algorithm that generates new tracking hypothesis for target estimation whenever new measurements arrive. The previous hypothesis are stored and used recursively on all new hypothesis calculations. As the number of measurements and targets increases, the number of hypotheses also increases, thus making the computation cost increase exponentially. Similar to JPDAF, the MHT has been adapted in order to be computationally tractable in \cite{streit1995probabilistic, willett2002pmht} at the cost of accuracy. Different from the late two approaches, the RFS combines the states of multiple targets into a set of single-target states. The computational complexity is also a key disadvantage of RFS. 

	
	
To address the computational complexity in MTT, the paper focuses on developing a new learning approach. We consider the scenario when a number of targets move along a known road network with numerous sensors randomly placed at unique known locations. The target's data is recorded when it passes by a sensor's location. It is assumed that sensors do not have physical capabilities to acquire target identification but only the velocity and time when a target passes by. The objective of this paper is to develop a learning-based approach that can effectively extract and associate correlated data to each target. More precisely, we seek to classify all data obtained from spatially distributed sensors into datasets and associate each of these sets to one target. Towards this objective, we first apply Deep Neural Network (DNN) and K-means to study their effectiveness. Motivated by the advantages of the two methods, we further develop a new multi-layer K-means (MLKM) algorithm that integrates the structure of both methods. Motivated by the backpropogation in deep learning methods, we also develop a new error correction operation in MLKM to further improve the accuracy. We then provide simulation examples and compare the performance of these methods on numerous datasets.
	
	The remainder of this paper is organized as follows. Section~\ref{sec:PF} provides a formal definition of the the problem to be investigated in this paper. Section~\ref{sec:alg} first describes the design of K-means and DNN for data association and then present the new MLKM algorithm. Section~\ref{sec:simu} provides numerous simulations that illustrate the performance of the proposed algorithms. Finally, Section~\ref{sec:con} provides a brief summary and discusses some potential future work.

	\section{Problem Formulation}\label{sec:PF}
	
	In the context of this paper, we consider a 1-dimensional road, having a length denoted as $D\in \mathbb{R}_{>0}$. Let $\mathcal{S}=\{S_1,\cdots,S_{N_S}\}$ be a set of $N_S\in \mathbb{Z}_{>0}$ sensors placed along the road at the locations $\mathcal{R}=\{R_1,\cdots,R_{N_S}\}$, where each element in $\mathcal{R}$ is unique and valued in the range $(0,D)$.
	
	As a target passes the sensor, the sensor collects the target's information. This information includes the velocity of the target (denoted as $v$) and a time stamp representing when the target passes the sensor (denoted as $t$). The information collected about the target is disassociated with the target, meaning that the target for which the information was received cannot be directly identified using the information. 
	
	Let $\mathcal{A}=\{A_1,\cdots,A_{N_A}\}$ represent a set of $N_A\in \mathbb{Z}_{>0}$ targets. It is assumed that each target passes each sensor exactly one time. The sensors store their collected information in the matrices $V\in \mathbb{R}^{N_{A}\times N_{S}}$ and $T\in \mathbb{R}^{N_{A}\times N_{S}}$. The $m$th velocity and time stamp measurements obtained by the $n$th sensor is recorded as $V_{m,n}\in V$ and $T_{m,n}\in T$, respectively (where $n\in \{1,\cdots,N_S\}$ and $m\in \{1,\cdots,N_A\}$). Let the information $\mathcal{X}\subset \mathcal{R}^1$ be the collection of velocity data $V$ and time stamp information $T$, organized such that each element $X_{n}^m\in \mathcal{X}$ is a set of data containing the values $V_{n,m}$ and $T_{n,m}$. More precisely, the set of all observations from the sensor network is represented as $\mathcal{X} = \{X_1^1, X_1^2,...,X_1^{N_A},...,X_{N_S}^{N_A}\}$.

	
	\begin{figure}[ht!]
		\centering
		\includegraphics[width=8cm]{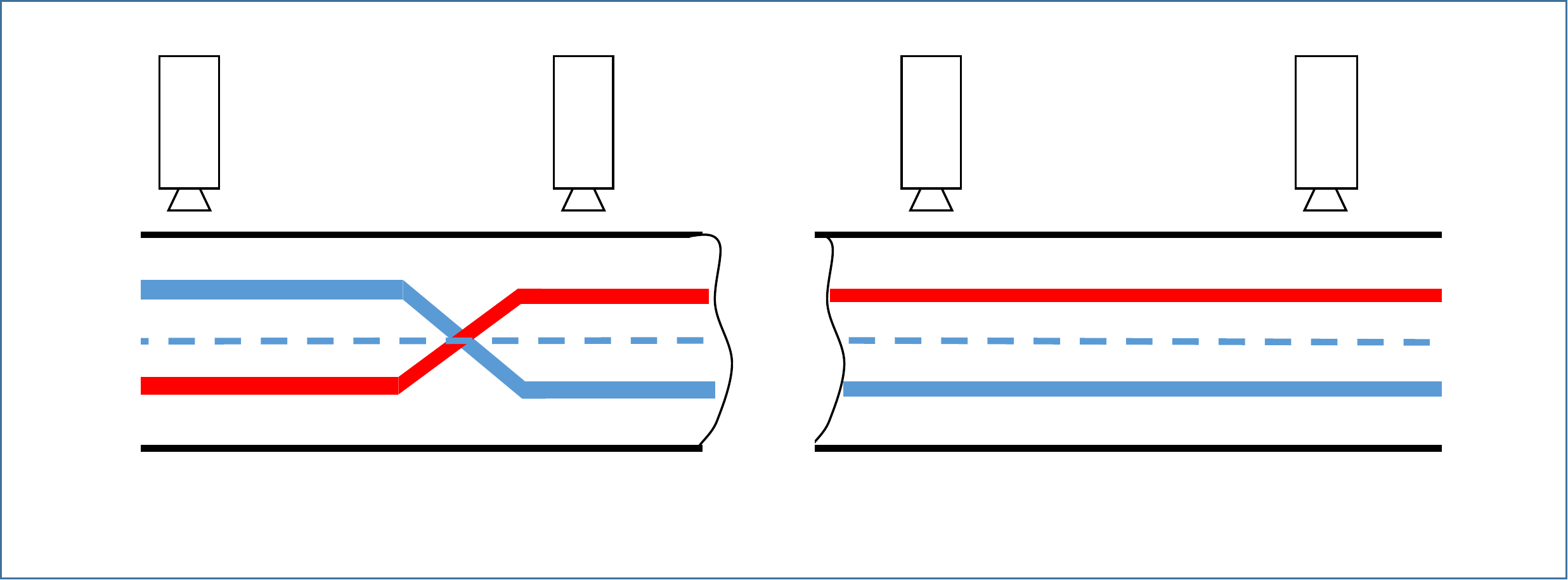}
		\caption{1-D road with multiple sensors.}
		\label{fig:roadNW}
	\end{figure}
	
	Let $\mathcal{X}_{ideal} \subset \mathcal{R}^2$ be the desired outcome of the form given by
	\begin{align}\label{eq:ideal}
	\mathcal{X}_{ideal}=
	\begin{Bmatrix}
	X_1^1&\cdots&X_{N_S}^1\\
	\vdots&\ddots&\vdots\\
	X_1^{N_A}&\cdots&X_{N_S}^{N_A}\\
	\end{Bmatrix},
	\end{align}
	where each row of $\mathcal{X}_{ideal}$ represents the measurements of all sensors regarding the same target. For example, the $i$th row of $\mathcal{X}_{ideal}$ is the dataset associated with target $A_i$ observed by different sensors at different time instances. Because the targets' velocities fluctuate as they move along the road network, the  sequence of targets observed by one sensor may not be the same as that observed by another one, leading a necessary data pattern recognition problem. More precisely, given a set of observations from all sensors as $\mathcal{X} = \{X_1^1, X_1^2,...,X_1^{N_A},...,X_{N_S}^{N_A}\}$,\footnote{In this case, $X_n^m$ means the $m^{th}$ sample collected by the $n^{th}$ sensor, ($m$ does not identify a specific target).} our goal is to classify the data $\mathcal{X}$ in the desired form of $\mathcal{X}_{ideal}$ as in~\eqref{eq:ideal}. For the simplicity of representation, we assume that no false alarm or missing detection will occur although the proposed methods can be potentially augmented appropriately to deal with false alarm and missing detection.
	
	To evaluate the performance of the proposed method, labeled data is needed to obtain the percentage of true association between targets and their measurements. Let the information $\mathcal{Y}$ be the collection of information that includes the true association of targets to their measurements. The information $\mathcal{Y}$ is structured similarly to $\mathcal{X}$ except that each element $Y_n^m \in \mathcal{Y}$ also includes the value $A_i$, which describes the target identification that the measurement corresponds to.
	
	Algorithm \ref{alg:data-gen} is the pseudo code that describes how multiple datasets of $\mathcal{Y}$ are generated. The data sets of $\mathcal{Y}$ are generated in iterations. Let $N_K$ be the total number of iterations and let $K$ be the current iteration step. In each iteration step, there is a sub-iterative process. The sub-iterative process generates and records the trajectory for a single target. The trajectory is generated by first defining the initial time and velocity values for the target (line 5 and 6 of Algorithm 1). These values are selected from a uniform random variable, where $(v_{min}, v_{max})$ and $(t_{min},t_{max})$ indicate, respectively, the upper and lower bound for initial velocity and time. {The function $g(\cdot,\cdot)$ (line 7 of Algorithm \ref{alg:data-gen}), which is not fully described here, represents how a unique trajectory is generated for the profile $V_i(t)$. The function $g(\cdot,\cdot)$ relates the previous velocity and time values to random variations in order to generate the current velocity profile.} Then, according to this trajectory the function $h(\cdot,\cdot,\cdot)$ is used to insert measurements into the appropriate elements of $\mathcal{Y}_K$ (line 8 of Algorithm 1). The velocity and time measurements associated with target $A_i$ are inserted into the element location of $\mathcal{Y}_K$ according to the sensor positions $\mathcal{R}$. After inserting all measurements, the information $\mathcal{Y}_K$ is saved (line 10 of Algorithm 1). 
	
	\begin{algorithm} 
		\caption{Data generation}\label{alg:data-gen} 
		\begin{algorithmic}[1] 
			\State{Output: Generate multiple $\mathcal{Y}$ from $N_S$ sensors on a road with length $D$} 
			\State $\mathcal{R} =  \{R_1,R_2,...,R_m\}$; 
			\For {$K =1:N_K$}   
			\For {$i =  1:N_A$} 
			\State  $v_{o_i}= P(x) = \mathcal{N}(\mu,\sigma^2) $; 
			
			\State  $t_{o_i} =t_{o_i-1} \pm \sim U (t_{min},t_{max})$; 
			
			\State $V_i(t) = g(v_{o_i},t_{o_i})$; 
			
			\State $\mathcal{Y}_K\leftarrow h(V_i(t),A_i,\mathcal{R})$ 
			\EndFor 
			\State Save $\mathcal{Y}_K$ 
			\EndFor 
		\end{algorithmic} 
	\end{algorithm}

	\section{Algorithm}\label{sec:alg}
	To solve the aforementioned problem, in this section a set of machine learning algorithms is derived to classify the information $\mathcal{X}$, such that each target-measurement pair can be identified. In addition, we are interested in conducting a comparison of the performance of these algorithms for different datasets. Using the appropriate machine learning techniques the random data $\mathcal{X}$ will be re-organized in order to achieve the mapping described as
	\begin{align}
	\Theta:(\mathcal{X} \subset \mathcal{R}^1) \mapsto (\mathcal{X}_{ideal} \subset \mathcal{R}^2),\nonumber 
	\end{align}
	where each row of $\mathcal{X}_{ideal}$ represents a set of all measurements associated to the $m$th target. The performance of the proposed machine learning algorithms will be evaluated by the associated accuracy levels over all datasets. We first apply both shallow machine learning algorithms, i.e., K-means and K-means++, and deep learning algorithms, i.e., Deep Neural Network (DNN), to tackle the data association problem. Then we develop a new Multi Layer K-means (MLKM) algorithm by leveraging the structural advances of both K-means and DNN. 
	
	
\subsection{K-means, K-means++, and Deep Neural Network}
	
K-means~\cite{kanungo2002efficient} and K-means++ \cite{arthur2007k} are perhaps the most common methods for data clustering. For a set of data points in a $N$-dimensional system, the two algorithms perform clustering by grouping the points that are closer to the optimally placed centroids. From the machine learning perspective, K-means learns where to optimally place a pre-defined number of centroids such that the cost function, defined as
\begin{align}
\Phi_Y(\mathcal{C}) = \sum_{y\in Y} d^2(y,\mathcal{C})\textup{,} \nonumber 
\end{align}
is minimized, where $d(y,\mathcal{C}) = {\min}_{y\in Y}\left \| y-c_i \right \|$ represents the distance between a sub-set of measurements $Y$ and a centroid $c_i$ and $\mathcal{C} = \{c_1,...,c_k\}$ represents the set of centroids
 The associated cost function is the sum of the Euclidean distances from all data points to their closer centroid. The cost function and optimization algorithms are the same for K-means and K-means++ while the only difference between them is that K-means++ places the initial guesses for the centroids in places that have data concentration. When the measurements (data points) are distributed along the time axis and velocity profiles are close (as is the case in the studied problem), the centroids tend to be placed in positions that cause data from different targets to be clustered together. This happens because the boundaries defined in K-means are limited by the 2-dimensional nature of the problem. 
 
 Taking into consideration the need for more appropriate boundaries, we also apply the Deep Neural Network (DNN)~\cite{hinton2002neural} in data association. It's well known that DNN is good at recognizing underlying patterns and defining better decision boundaries and hence can potentially provide better performance for the data association problem. However, DNN demands labeled datasets since it is a supervised learning method. For the purpose of evaluating DNN capabilities, a slight modification of the problem is considered. Instead of a complete unlabeled dataset $\mathcal{X}$, part of the measurements is already labeled appropriately. As described in section \ref{sec:PF}, the data available at each measurement is composed of position $p$, time $t$ and velocity $v$. To provide extra features for DNN to learn the data pattern, we also add extra features $vt$, $v^2t$, and $vt^2$. Defining and training a DNN require the definition of a framework consisting of: {i - cost function}; {ii - activation function}; {iii - optimizer}; {iv - number of neurons per layer}; and, {v - number of hidden layers}. Table \ref{tb:parameters_nn} presents the parameters used in the framework.
	
	\begin{table}[h]
		\centering
		\caption{DNN configuration parameters.}
		\begin{tabular}{|l | c|}
			
			\hline 
			Framework & Definition \\ 
			\hline 
			Cost Function & Softmax \cite{bishop2006pattern} \\ 
			
			Activation Function & Relu \cite{glorot2011deep} \\ 
			
			Optimizer & Adam Optimizer \cite{kingma2014adam} \\ 
			
			Number of Neurons &  8\\ 
			
			Number of Layers&  2\\ 
			\hline 
		\end{tabular}
		\label{tb:parameters_nn}%
	\end{table}   
	
The selection of the cost, activation, and optimization functions follows the standard choices for classification problems as in \cite{sun2016scene, zhou2016polarimetric}. In addition, selecting the number of hidden layers and neurons demand some empirical tests. For DNN with a high number of hidden layers and neurons, the use of a small dataset increases the chance of overly fitting the training data, resulting in low accuracy on the test data. In order to avoid the overfitting problem, the DNN is built with 2 hidden layers composed of 8 neurons each so that the number of parameters to be trained on the network is smaller. These parameters are tested to show the best accuracy for the studied problem.

\subsection{K-means with data pre-processing}\label{sub:pre-processing}

\begin{figure}[h!]
		\centering
		\includegraphics[width=7cm]{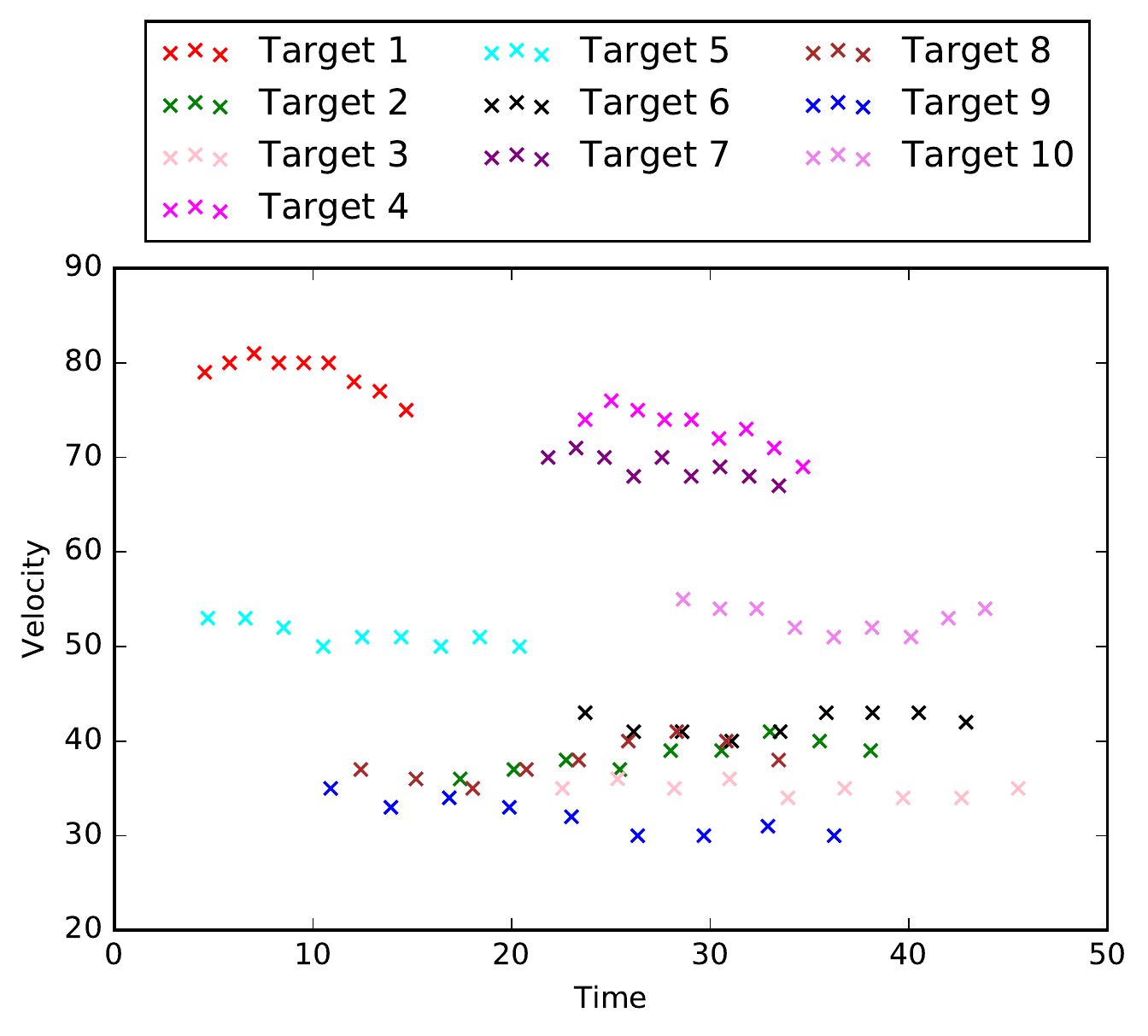}
		\caption{Dataset without pre-processing.}
		\label{fig:thirdDataset}
	\end{figure}

Extracting data association from the raw data $\mathcal{X}$ using the K-means algorithm can be challenging when different targets have similar velocity profiles because K-means seeks to classify targets with similar velocity and time combination. Without considering the locations of different sensors, the data association using K-means algorithms is expected to yield low performance. To obtain more accurate data association, an appropriate preprocessing of $\mathcal{X}$ is needed. Figure~\ref{fig:thirdDataset} show the case where overlapping between measurements in $\mathcal{X}$ occur without any preprocessing of the data.
	
	The method for preprocessing the data is to project the  measurements in $\mathcal{X}$ and estimate what would be that target's measurement obtained by sensor $S_i$. A notation for such a mapping is given by
	\begin{align}
	&\mathcal{F}:(S_n \in \mathcal{S}) \mapsto (\mathcal{S}_i \in \mathcal{S}),\forall~ 0\leq n\leq N_S,~0\leq m\leq N_A.\nonumber 
	\end{align}
The data obtained after pre-processing is represented as
	\begin{equation} \label{eq:afterPreprocessing}
	\mathcal{X'} = \begin{Bmatrix}
	{X'}_1^1&\cdots&{X'}_{N_S}^1\\
	\vdots&\ddots&\vdots\\
	{X'}_1^{N_A}&\cdots&{X'}_{N_S}^{N_A}\\
	\end{Bmatrix}.
	\end{equation}
In this case, the sensor $S_i$ (the $i$th sensor) is considered as a reference. The other sensor readings will be approximately transferred to estimated readings by the sensor $S_i$. For each sensor $S_{n}$, velocity $V_{n,m}$, time $T_{n,m}$, and position $R_n$ information will be used to calculate the approximate time instant ${T'}_{n,m}$ that the target passed sensor $S_i$. Both velocity $(V_{n,m})$ and the new time ${T'}_{n,m}$ will form the data ${X'}_n^m$ after pre-processing. 
	
\begin{figure}[h!]
		\centering
		\includegraphics[width=7cm]{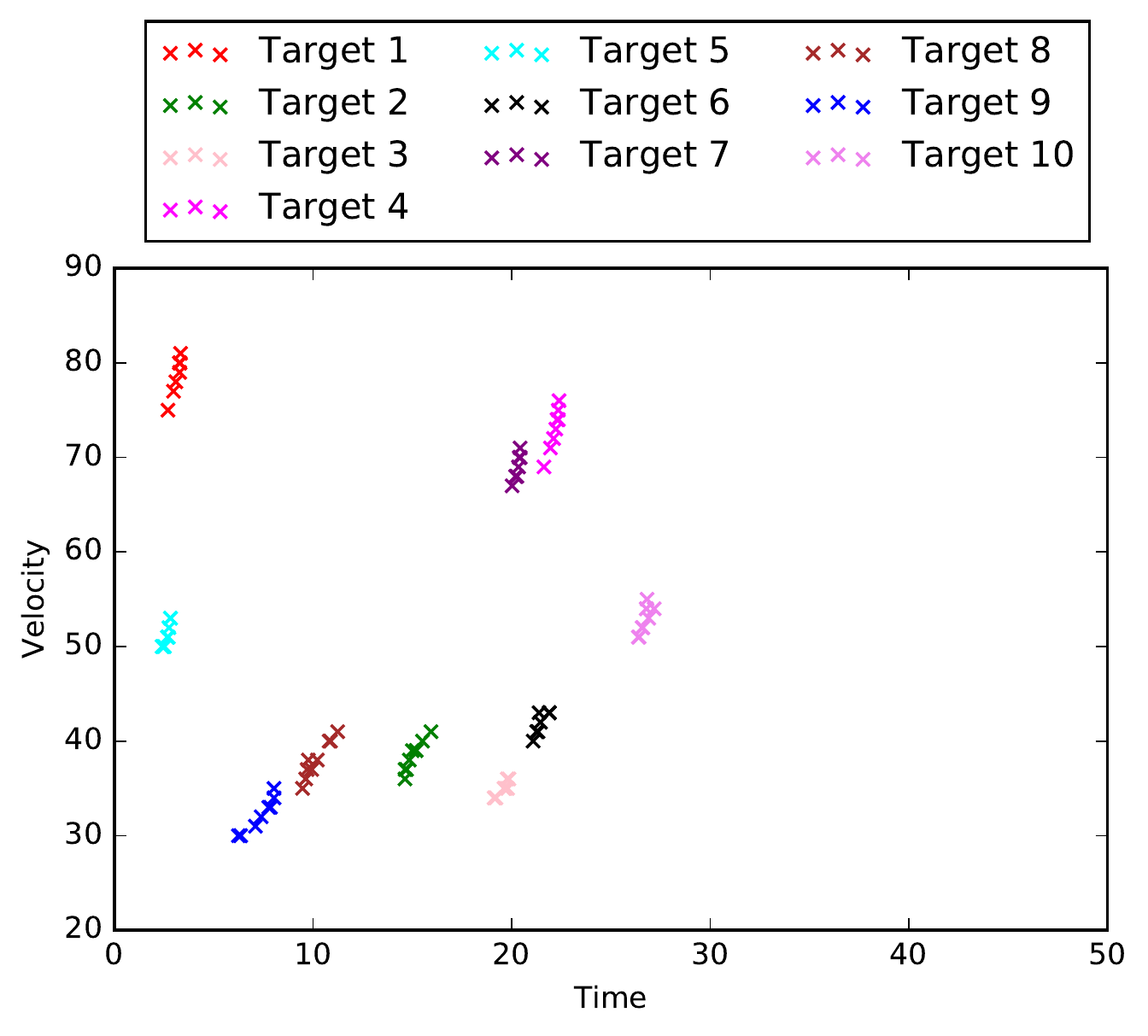}
		\caption{Dataset in Figure \ref{fig:thirdDataset} after pre-processing.}
		\label{fig:ProcessedDataset3}
	\end{figure}

The definition of this pre-processing is given by
	\begin{align}  
	{T'}_{n,m} =T_{n,m} ~-~ \frac{(R_n-R_i)}{V_{n,m}}, \nonumber
	\end{align}
and describes the backward extrapolation of the time from the $n$th sensor to sensor $S_i$. For example, the 4th measurement obtained by sensor $S_2$ includes $V_{2,4}$ = 20 m/s and $T_{2,4}$ = 50s. Let the distance between sensor $S_1$ and $S_2$ (evaluated as $R_2 - R_1$) be 100 units. The estimated time associated with the measurement $X_2^4$ is then calculated as ${T'}_{n,m}$ = 45s. Conducting the same pre-processing strategy to the information $\mathcal{X}$ results in the converted time estimation of all measurements with respect to sensor $S_1$. Figure \ref{fig:ProcessedDataset3} show the plot of the datasets used in Figure~\ref{fig:thirdDataset} respectively after applying the proposed data pre-processing method. As can be observed from Figure~\ref{fig:ProcessedDataset3}, the processed data yields some patterns that are more likely to be recognized by machine learning algorithms, such as K-means.
	

	When the dataset is preprocessed and projected backwards with reference to sensor $S_i$, the obtained dataset is given by $\mathcal{X'}$ in~\eqref{eq:afterPreprocessing}, where ${X'}_n^m$ consists of information $V_{n,m}$ and ${t'}_{n,m}$ that correspond to sensor $S_n$ located at $R_n$. Therefore, the original time $t_{n,m} $ can be approximately obtained via the following operation:
	\begin{equation*}
	t_{n,m} = {t'}_{n,m} + \frac{R_n-R_i}{V_{n,m}}.
	\end{equation*}
	
Algorithm \ref{alg:K-means} shows the pseudo code to solve the data association problem by using K-means with pre-processing. 	Although the K-means++ algorithm provides high accuracy for small datasets, its accuracy, however, decreases significantly for large datasets due to the inevitable overlaps among a higher number of targets and sensors even when pre-processing is performed (c.f. the simulation examples in Section~\ref{sec:simu}). To further solve the issue, we next propose a new multi-layer K-means (MLKM) algorithm that leverages the advantages of both K-means and DNN.
	
	\begin{algorithm}[h!]
		\caption{K-means with Preprocessing Algorithm}\label{alg:K-means}
		\begin{algorithmic}[1]
			\State{Input: $ \mathcal{X} $}
			\State{Output: Obtain ~$\mathcal{X}'' = \begin{bmatrix}
				{X}_1^1 &{X}_2^1&\ldots&{X}_m^1\\
				\vdots&\vdots&\vdots&\vdots\\
				{X}_1^n &{X}_2^n&\ldots&{X}_m^n\\
				\end{bmatrix}$}
			\While{not at the end of the file} 
			\State
			\State $X[m \times n] \leftarrow \begin{bmatrix}
			V_1^1 &t_1^1  &P_1 \\ 
			\vdots& \vdots &\vdots \\ 
			V_1^n&t_1^n  & P_m
			\end{bmatrix}$;
			\EndWhile
			\While {not at the end of $X[m\times n$]}
			\State $tp_i^j = t_i^j - \frac{P_i- P_1}{V_i^j};$
			\State $\mathcal{X'}$ $\leftarrow V_i^j, {t'}_i^j, P_i;$
			\EndWhile
			\State $\mathcal{C} \leftarrow \{c_1,c_2,...,c_n\}$
			\While {$\Delta \Phi_Y(\mathcal{C}) \geq \varepsilon$}
			\For {$i\leq m*n $}
			\State $y_i = arg \min_{j} \left \|\mathcal{X}_i'-c_j  \right \|^2 $
			\EndFor 
			\For {$j\leq m $}
			\State $c_j =\frac{\sum_{y\in c_j}\mathcal{X}_i'}{|c_j|}$
			\EndFor 
			\EndWhile
			\While {not at the end of $\mathcal{X'}$}
			\State $t_i^j = {t'}_i^j + \frac{P_i-P_1}{V_i^j};$
			\State $\mathcal{X}'' \leftarrow V_i^j, {t}_i^j, P_i;$
			\EndWhile
			\State return $\mathcal{X}''$
		\end{algorithmic}
	\end{algorithm}
	
\subsection{Multi Layer K-means++}
	
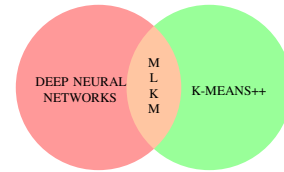
\begin{figure}[h!]
		\centering
		\begin{tikzpicture}
		
		\begin{scope}[blend group = soft light]
		
		\fill[red!40!white] (210:0.85) circle (1.1);
		\fill[green!40!white]  (330:0.85) circle (1.1);
		\end{scope}
		
		\node [font=\tiny] at ( 200:1.03)   {DEEP NEURAL};
		\node [font=\tiny] at ( 210:1.13)   {NETWORKS};
		\node [font=\tiny] at ( 335:1.1)   {K-MEANS++};
		\node [font=\tiny] at (270:0.10) {M};
		\node [font=\tiny] at (270:0.30) {L};
		\node [font=\tiny] at (270:0.50) {K};
		\node [font=\tiny] at (270:0.70) {M};
		\end{tikzpicture} 
		\caption{MLKM integrates K-means++ and Neural Networks.}
		\label{fig:MLKM}
	\end{figure}

	
	K-means++ provides highly accurate data association when the dataset is small, but its performance drops drastically when the dataset starts to increase. In contrast, Deep Neural Networks (DNN), equipped with a multi-layer structure and the error backpropagation philosophy provides unprecedented accuracy for big datasets but demands very large labeled datasets for training. In order to solve the data association problem with a large number of targets and sensors, we proposed a multi-layer K-means (MLKM) method, which integrates the DNN's multi-layer structure with the clustering capabilities of K-means++. Figure~\ref{fig:MLKM} shows the relationship between MLKM and DNN/K-means.	

\begin{figure}[h!]
		\centering
		\includegraphics[width=8cm]{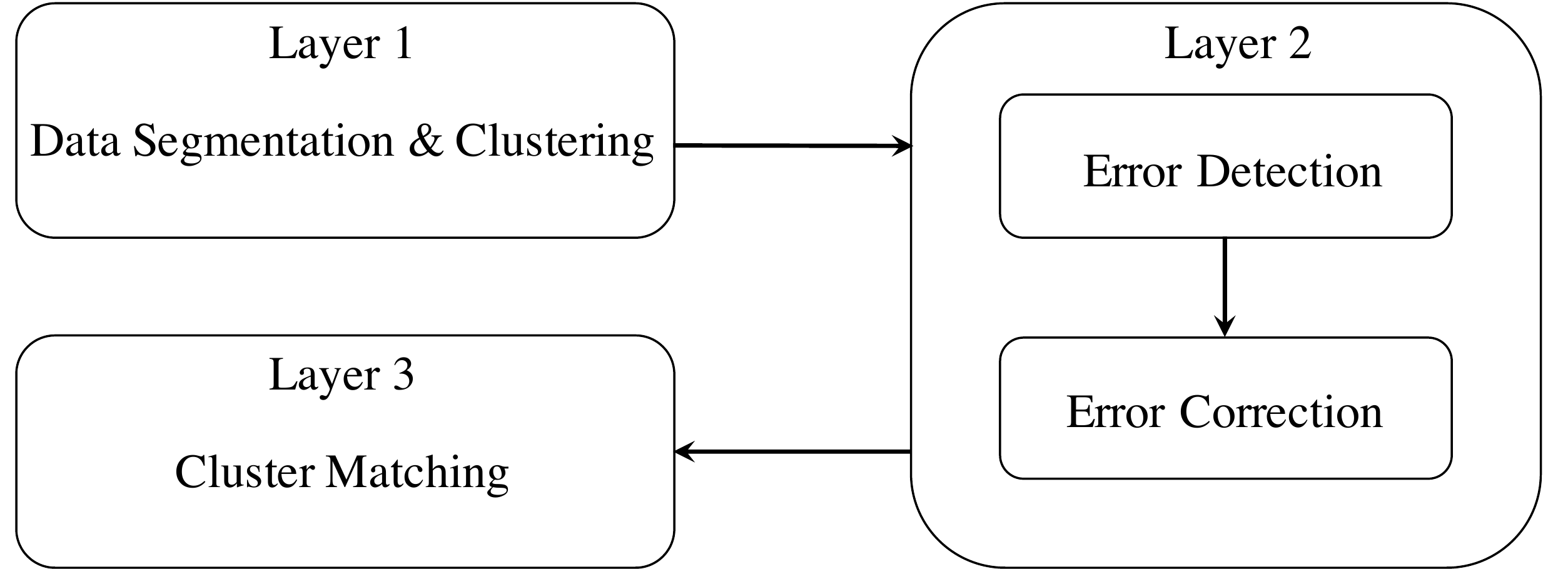}
		\caption{MLKM algorithm flow diagram.}
		\label{fig:MLKMdiagram}
	\end{figure}

The current MLKM algorithm is performed via 3 layers: \textit{(i) data segmentation and clustering} - Segment the data in smaller groups with the objective of improving the first K-means++ execution over the measurements; \textit{(ii) - Error detection and correction} - Analyze the clustered data by searching for errors through predefined rules and re-cluster the data using Nearest Neighbor concepts. Note that the K-means++ associates the data closer to the optimally placed centroid by taking into consideration only the Euclidean distance between data point and centroid; and, \textit{(iii) - Cluster matching} - match the clusters of each segment by referring the cluster centroids of all segments to the cluster centroid of the first segment and grouping them based on K-means++. Figure \ref{fig:MLKMdiagram} shows the flowchart of the proposed MLKM algorithm.
	
	The MLKM algorithm assumes that no false alarms or missing detection will occur. We further assume that there are $K$ sensors per segment without loss of generality. 
	
\subsubsection*{Layer 1 \textit{(Data Segmentation \& Clustering)}}
	
The dataset $\mathcal{X} = \{X_1,X_2,\cdots, X_n\}$, composed of $N_s$ vectors (measurements from all sensors and $m$ targets), is divided into $E$ segments, where 
\begin{align*}
E = \left\{
\begin{array} {ll}
N_s/K,&N_s\%K=0,\\
N_s/K+1,&\text{otherwise}.
\end{array}\right.
\end{align*}
In other words, when $N_s\%K\neq 0$, the last segment will contain measurements from less than $K$ sensors. Without loss of generality, we assume that $N_s\%K=0$ in the following of this paper. Accordingly, each $\mathcal{X}_e = \{X_{(e-1)K+1},\cdots, X_{eK}\}$ subset of $\mathcal{X}$ is defined as 
	\begin{align}
	\mathcal{X}_e= \bigcup_{j=(e-1)K+1}^{eK} X_j \textup{.}\nonumber
	\end{align}
First, K-means++ algorithm is applied to each $\mathcal{X}_e$ without using pre-processing. Here the preprocessing is not used because conducting preprocessing for a large number of sensors can lead to large errors since the velocity of each target may change significantly from the first sensor to the last sensor. Then by aggregating the centroids of all $N_s/K$ groups, we can obtain a set of centroids, defined as $\mathcal{C}'_e = \{c_1^e,c_2^e,\cdots,c_m^e\}$, and a set of measurements, defined as $\mathcal{Y}_m^e$, associated with each $c_m^e$ centroid. For both $c_m^e$ and $\mathcal{Y}_m^e$, $m$ refers to the cluster centroid and $e$ refers to the segment. 
	
\subsubsection*{Layer 2 \textit{(Error Detection \& Correction)}}
The first layer seeks to roughly associate data for each group. Because of the potential data association error, we here propose to perform error detection and correction, which is motivated by the backpropogation operation in DNN. The error detection is to use logic rules to determine if wrong data association appears in $\mathcal{Y}_m^e$. The error correction will conduct data re-association from all $\mathcal{Y}_m^e$ when at least one data in $\mathcal{Y}_m^e$ is labeled incorrectly.

We first start with the error detection. The rules for error detection include:
	\begin{itemize}
		\item $|\mathcal{Y}_m^e| \neq m$;
		\item $\forall i \neq j$, $X_i \in \mathcal{Y}_m^e$, $X_j \in \mathcal{Y}_m^e$, $p_i=p_j$; and,
		\item $\forall i \geq j$, $X_i \in \mathcal{Y}_m^e$, $X_j \in \mathcal{Y}_m^e$, $t_i > t_j$;
	\end{itemize}
where $|\mathcal{Y}_m^e|$ indicates the cardinality of $\mathcal{Y}_m^e$. The first rule means that less or more than $m$ targets are the points in $\mathcal{Y}_m^e$ assigned. Error can be determined in this case because no target will be recorded exactly once by one sensor. The second rule means that more than one sensory measurements were obtained by one sensor. The third rule means that target was recorded later at the previous sensor, which cannot happen because it always takes some time for a target to move from the previous sensor to the next one. If any element of $\mathcal{Y}_m^e$ satisfies any of the three rules, the entire measurements from the same sensor is considered to contain erroneous data association and will be stored in $\mathcal{Y}^{e*}$, where $\mathcal{Y}^{e*}$ refers to the wrong data association in $\mathcal{X}_e$. 

We now describe the error correction. Because $\mathcal{Y}^{e*}$ is the subset of $\mathcal{X}_e$ that needs to be re-classified, it is necessary to re-associate data such that the rules will not hold in the new data clustering. To achieve this, we propose to use a nearest neighbor approach. Specifically, an element of $\mathcal{Y}^{e*}$, labeled as $(t_\ell, v_\ell)$, for sensor at $R_\ell$, is evaluated against every measurement in the next sensor, \textit{i.e.,} sensor at $R_{\ell+1}$, to obtain the best match. The evaluation is accomplished via the following optimization problem:
	\begin{align}
	\arg \min_{\kappa} ~& t_{\ell+1}-(t_\ell + \frac{\left \| R^\kappa_{\ell+1}-R_{\ell} \right \|}{v_\ell}), \nonumber\\
	~\textup{s.t.} ~ &\kappa \in \{1,2,\cdots,\abs{X_{\ell+1}}\}. \nonumber 
	\end{align}
With this procedure, all $\mathcal{Y}^{e}$ is updated with the corrected clusters and all $c_m^e$ are re-calculated based on the new $\mathcal{Y}^{e}$. The new corrected set of centroids is updated for all segments and grouped into $\mathcal{C} =\{\mathcal{C}_1~\mathcal{C}_2~...~\mathcal{C}_E \}$.
	
	\subsubsection*{Layer 3 \textit{(Cluster Matching)}}
	
	From layers 1 and 2 we have correlated data from each segment. However, it is still unclear how to associate the clusters among different segments according to the specific targets. In Layer 3 we solve the ``Cluster Matching" problem. Specifically, all $\mathcal{C}^e$ is projected to $\mathcal{C}^1$ using the pre-processing technique (each element in $\mathcal{C}^e$ is $c_m = \{v_m,t_m\}$). Then K-means++ is applied to the projected centroids to find the correct association among the clusters. Since we have solved data association within each segment and cluster association among different segments, now we can reorganize the data as proposed in \eqref{eq:ideal}.

	

	\section{Simulation}\label{sec:simu}
	
	In order to demonstrate the performance of the proposed MLKM algorithm, a single lane road scenario has been implemented with $N_A$ targets and $N_S$ sensors placed along the road. The sensors are capable of acquiring information about time, velocity, and position. In addition, each sensor is unable to detect the target ID. 
	
	To evaluate the performance of MLKM, simulations provide results with respect to the four machine learning algorithms: K-means, K-means++, DNN, and MLKM. The algorithms are applied to 4 different datasets. Statistical information regarding the results for these different datasets are provided based on the data association \textit{accuracy}. In this paper, accuracy is defined as the ratio between correctly classified data ($M_{cr}$) and the total number of data ($M_{t}$) in the form of $\frac{M_{cr}}{M_{t}}\times100\%$. The statistical information provides the average (middle - orange bar), maximum (right - gray bar) and minimum (left - blue bar) accuracy obtained for each of the learning algorithms, as shown in Figures \ref{fig:kmeanscomp10t10s}, \ref{fig:kmeanscomp50t20s} and \ref{fig:ModelComp}.

	\subsubsection{K-means and K-means++} 
\begin{figure}[h!]
		\centering
		\includegraphics[width=7.5cm]{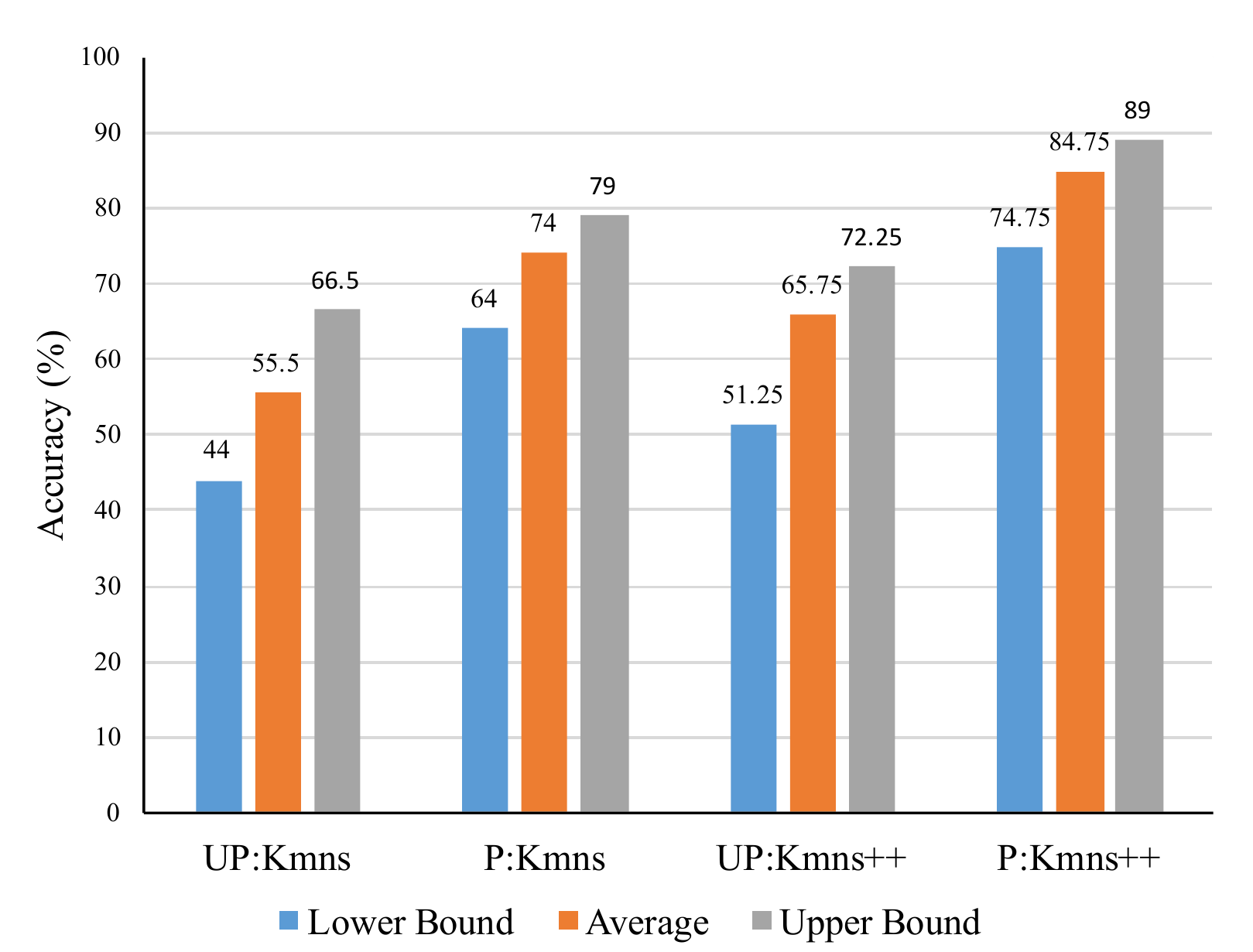}
		\caption{K-means and K-means++ accuracy for Simulation 1 parameters on unprocessed (UP) and preprocessed (P) data.}
		\label{fig:kmeanscomp10t10s}
	\end{figure}

\begin{figure}[h!]
		\raggedleft
		\includegraphics[width=8cm]{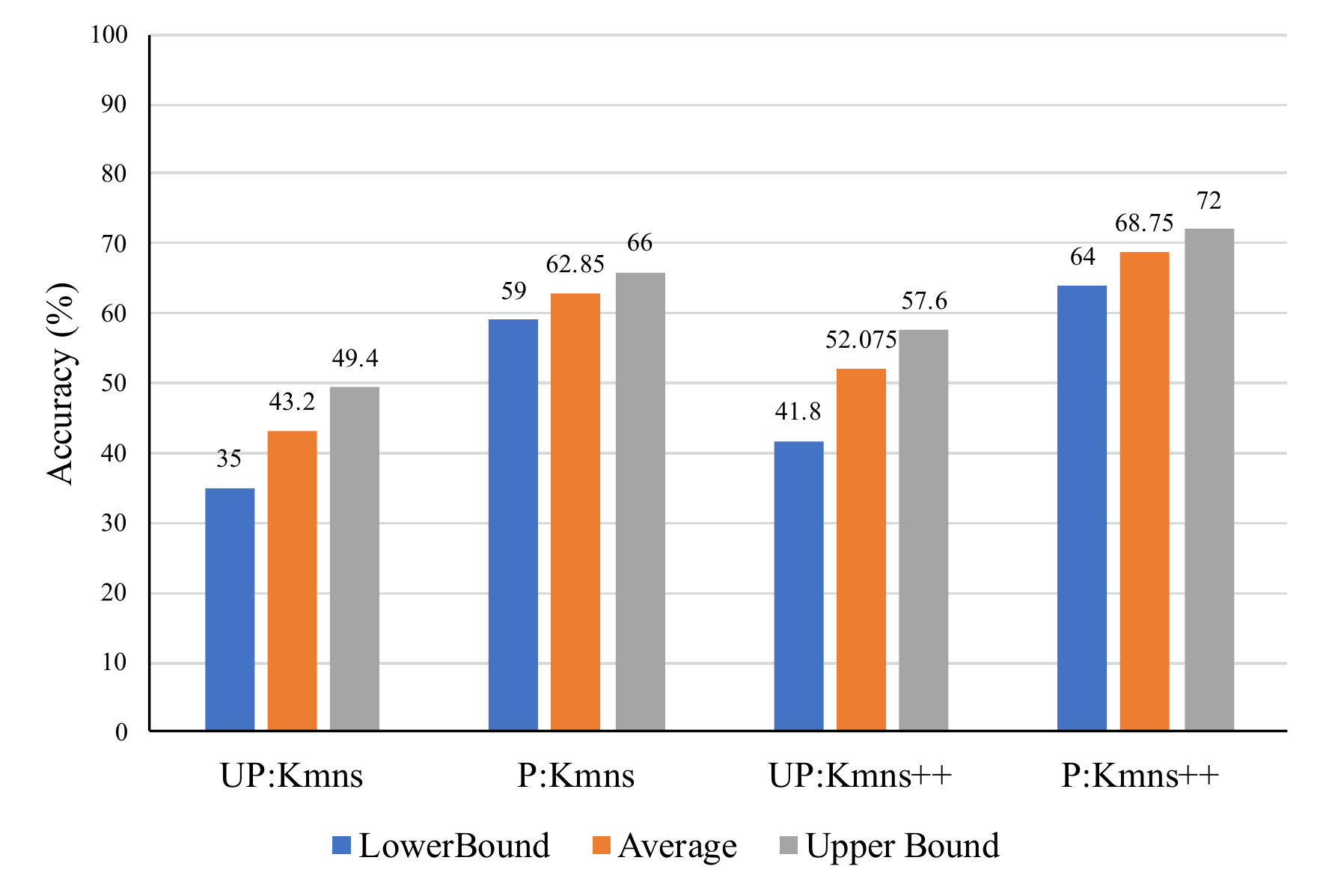}
		\caption{K-means and K-means++ accuracy for Simulation 2 parameters on unprocessed (UP) and preprocessed (P) data.}
		\label{fig:kmeanscomp50t20s}
	\end{figure}

The first set of simulations evaluate the performance of K-means and K-means++ based on two criteria: (i) unprocessed vs. preprocessed data, and (ii) using different values of $N_A$ and $N_S$. When the values of $N_A$ and $N_S$ increase, more data points are introduced into the dataset, leading to more overlapping among these data points. Figures \ref{fig:kmeanscomp10t10s} and \ref{fig:kmeanscomp50t20s} show the performance of K-means and K-means++ using the parameters listed in Table \ref{tb:sim_parameters}.  
	
	\begin{table}
		\centering
		\caption{Simulation parameters.}
		\begin{tabular}{|c|c|c|}
			
			\hline 
			& Simulation 1 & Simulation 2 \\ 
			\hline 
			$N_A$ & 10 & 50 \\ 
			\hline 
			$N_S$ & 10 & 20 \\ 
			\hline 
			$(V_{min},V_{max})$ & $\mathcal{N}(50,40)$ & $\mathcal{N}(50,40)$   \\ 
			\hline 
			$(t_{min},t_{max})$ & $\mathcal{U}(-10,30)$ & $\mathcal{U}(-10,30)$  \\ 
			\hline 
		\end{tabular} \label{tb:sim_parameters}
	\end{table}
	
	As can be observed, a higher accuracy is achieved using the preprocessed data than that using the unprocessed data. This can be seen by comparing average, maximum and minimum accuracy for the two methods that use the preprocessed data versus the the unprocessed data, as shown in Figure \ref{fig:kmeanscomp10t10s}. Using the raw data, the measurements associated with a specific target are sparse along the time axis. However, the velocity measurements from the same sensor are closely grouped along the velocity axis. These conditions contribute to incorrect clustering of the data. The preprocessing technique reduces the distance between target related measurements, therefore reducing the effect of the velocity measurements on the clustering. 
	
	A low accuracy is obtained for large values of $N_A$ and $N_S$. This can be observed by comparing average, maximum and minimum accuracy for different $N_A$ and $N_S$, as shown in Figures \ref{fig:kmeanscomp10t10s} and \ref{fig:kmeanscomp50t20s}. Similar to the unprocessed data, a large number of sensors/targets increases the density of measurement points. The concentration of measurements increases the probability that K-means/K-means++ clusters the data incorrectly (even with preprocessing).

	\subsubsection{DNN} The K-means/K-means++ fails to correctly cluster data when overlapping of measurements occurs. Deep neural networks (DNN) is used as an alternative approach because it has shown to provide unprecedented results to uncover patterns for large dataset classification. One necessary condition for DNN is the availability of labeled datasets for training. To meet the requirements of DNN, it is assumed that labeled data is available for training.

	The results for DNN are obtained using $N_A= 50$ targets and $N_S=50$ sensors. Assuming that a portion of the data association has already been identified, the objective is to train a neural network to label the unidentified measurements. The number of `training' sensors that provide labeled information and `testing' sensors that provide unlabeled information are provided in Table \ref{tb:accuracy_nn}. The accuracy is obtained for various proportions of `training' sensors to `testing' sensors. Table \ref{tb:accuracy_nn} also shows the accuracy obtained for different dataset configuration.

	\begin{table}[h]
		\centering
		\caption{DNN with different training and testing datasets.}
		\begin{tabular}{|c|c|c|c|}
			\hline 
			Train Sensors & Test Sensors & Train Accuracy & Test Accuracy \\ 
			\hline 
			20 & 30 & 98\% & 68\% \\ 
			\hline 
			25 & 25 & 97.8\% & 68\% \\ 
			\hline 
			30 & 20 & 99\% & 72\% \\ 
			\hline 
			40 & 10 & 98.6\% & 84.4\% \\ 
			\hline 
			45 & 5 & 98.9\% & 91.6\% \\ 
			\hline 
		\end{tabular} 
		\label{tb:accuracy_nn}%
	\end{table}
	
It can be observed that the training (respectively, testing) accuracy is high (respectively, low), when the testing dataset is relatively small. However, when the testing dataset is relatively high, the testing performance increases significantly (up to 91\%). A high training accuracy with a low testing accuracy means that DNN suffers from overfitting due to the small size of training dataset. Given this comparison, DNN is applicable when a large portion of training dataset is available to train the network for classifying a relatively small amount of measurements.
	
	\subsubsection{MLKM} K-means++ does not provide good accuracy for a high number of measurements but performs well when clustering small amounts of data. DNN can cluster large datasets but requires a large training dataset. MLKM combines the multi-layer back-propagation error correction from DNN and the clustering capabilities of K-means++. The DNN-inspired error correction significantly improves the performance of MLKM by preventing the clustering errors in layer 1 to propagate to the cluster association in layer 3. 
	
\begin{figure}[h!]
		\centering
		\includegraphics[width=8cm]{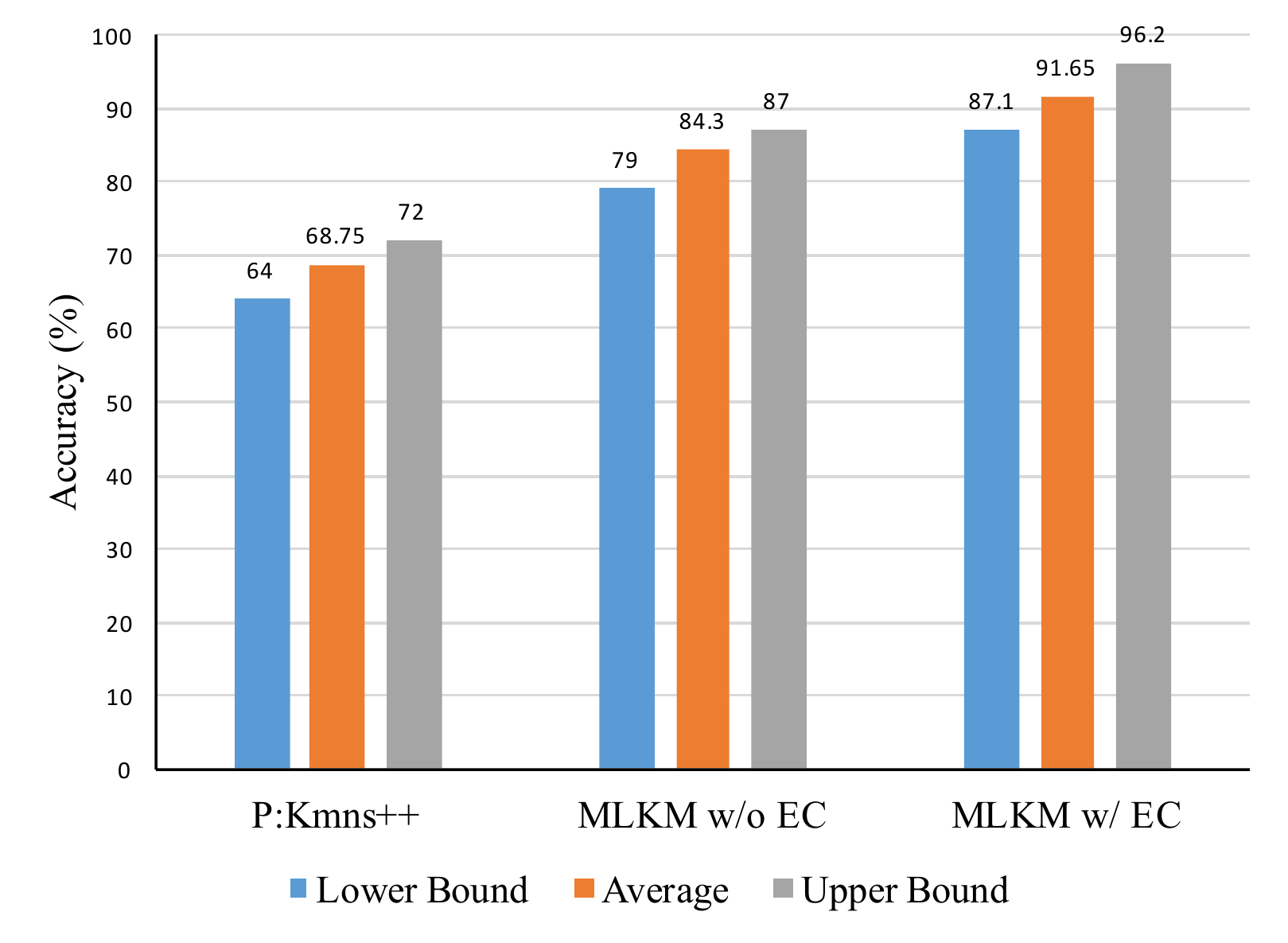}
		\caption{K-means++ for preprocessed data (P:K-means++), MLKM without error (MLKM w/o EC) correction and MLKM with error correction (MLKM w/ EC).}
		\label{fig:ModelComp}
	\end{figure}

	The results for the MLKM method are obtained using $N_A= 50$ number of targets and $N_S=20$ number of sensors. In addition, the time and velocity parameters are set to $(t_{min},t_{max}) = \mathcal{U}(-10,30)$ and $(V_{min},V_{max}) = \mathcal{N}(50,40)$, repectively. Figure \ref{fig:ModelComp} shows the performance of the MLKM method with and without error correction, as well as results using the standard K-means++ method.

It can be observed that a higher accuracy is achieved using MLKM than that using K-means++. Figure \ref{fig:ModelComp} shows the average, maximum and minimum accuracy for both methods. The error correction performed in layer 2 improves the average accuracy of MLKM by approximately 7\% (MLKM w/ EC 91.65\%; MLKM w/o EC 84.3\%). Even without the error correction, the MLKM method still outperforms the standard K-means++ algorithm.

	\section{Conclusion and Future Work}\label{sec:con}
	
	Identifying correct data-target pairing is essential for providing the situational awareness that the intelligent operation of unmanned systems heavily relies on. This paper studied data pattern recognition for multi-target localization from a set of spatially distributed sensors. In contrast to most existing methods that rely on probabilistic hypothesis estimation, we proposed to analyze the data correlation of unlabeled data from different sensors based on machine learning algorithms. Four algorithms, namely, K-means, K-means++, deep neural network, and MLKM, were used to solve the data association problem with different number of targets and sensors. In particular, the newly developed MLKM algorithm was shown to provide the best performance via leveraging the benefits of both K-means and deep neural networks.  In addition, simulation studies were provided to show that the proposed MLKM method offers a highly accurate solution for recognizing data patterns. Our future work will focus on studying more general road networks as well as the consideration of false alarms and missing detections.
	
	\bibliographystyle{IEEEtran}
	\bibliography{refs}
	
\end{document}